\documentclass[journal=nalefd,manuscript=letter]{achemso}
\usepackage{chemformula} 
\usepackage[T1]{fontenc}
\usepackage{textcomp}
\usepackage{amssymb}
\usepackage{amsmath}
\usepackage[utf8]{inputenc}

\usepackage{newfloat}
\DeclareFloatingEnvironment[name={Fig. S}]{suppfigure}

\usepackage{amsfonts}
\usepackage{mathbbol}
\usepackage{isomath}
\usepackage{mathptmx}

\usepackage{changes}

\author{Vahid Nasirimarekani}
\affiliation{Max Planck Institute for Dynamics and Self-Organization (MPIDS), 37077 G\"ottingen, Germany}
\author{Olinka Ram\`irez-Soto}
\affiliation{Max Planck Institute for Dynamics and Self-Organization (MPIDS), 37077 G\"ottingen, Germany}
\author{Stefan Karpitschka}
\affiliation{Max Planck Institute for Dynamics and Self-Organization (MPIDS), 37077 G\"ottingen, Germany}
\alsoaffiliation{Fachbereich Physik, Universit\"at Konstanz, 78464 Konstanz, Germany}

\author{Isabella Guido}
\email{isabella.guido@ds.mpg.de}
\affiliation{Max Planck Institute for Dynamics and Self-Organization (MPIDS), 37077 G\"ottingen, Germany}
\alsoaffiliation{Department of Physics, University of Surrey, Guildford, Surrey GU2 7XH, United Kingdom}

\title[An \textsf{achemso} demo]
  {Pattern formation under mechanical stress in active biological networks confined inside evaporating droplets}

\abbreviations{IR,NMR,UV}

\begin{document}


\begin{abstract}
	Active networks made of biopolymers and motor proteins are valuable bioinspired systems that have been used in the last decades to study the cytoskeleton and its self-organization under mechanical stimulation. Different techniques are available to apply external mechanical cues to such structures. However, they often require setups that hardly mimic the biological environment. In our study we use an evaporating sessile multi-component droplet to confine and mechanically stimulate our active network made of microtubules and kinesin motor proteins. Due to the well-characterized flow field inside an evaporating droplet, we can fathom the coupling of the intrinsic activity of the biological material with the shear stress generated by the flow inside the droplet. We observe the emergence of a dynamic pattern due to this combination of forces that vary during the evaporation period. We delineate the role that the composition of the aqueous environment and the nature of the substrate play in pattern formation. We demonstrate that evaporating droplets may serve as bioreactors that supports cellular processes and allows investigation on the dynamics of membraneless compartments. Such a setup is an original tool for biological structures to understand the mechanisms underlying the activity of the cytoskeleton under stress and, on the other hand, to investigate the potential of such adaptive materials compared to conventional materials.
\end{abstract}

\section*{Introduction}
Cytoskeletal assemblies in the form of networks of microtubules drive vital cellular processes such as intracellular cargo transport\cite{franker2013microtubule}, guide of cells during cellular migration\cite{small2002microtubules, watanabe2005regulation}, mechanical stability\cite{brouhard2018microtubule}, force generation, cell morphology \cite{kelliher2019microtubule} and division\cite{scholey2003cell}.
These functions are the result of the self-organization of microtubules and motor proteins that, by interacting at the molecular scales, determine the large-scale emergent behavior of the system \cite{sanchez2012spontaneous}.
Examples of such self-organization both in vivo and in vitro have been shown in several studies in the last decades, both experimentally and theoretically \cite{Alper, nedelec1997self, surrey2001physical, sanchez2011cilia, sanchez2012spontaneous, Goldstein, MONTEITH2016}.
Particularly interesting in nature is the influence of the external environment on such cytoskeletal networks and, especially, how they respond to external cues of mechanical nature. 
Despite their persistence length on the order of millimeters, highly curved microtubules can be observed in cells, suggesting that they experience large forces within the cytoplasmic space. Specifically, it was shown that they can bear mechanical loads that are transmitted over long distances inside the cytoplasm \cite{Wang2001,wang2002}.
Under external stimulation, also the activity of the motors can be influenced by the applied stress. The spatio-temporal distribution of the intrinsic driving forces which lead to the resultant emergent behaviour of the entire system can be tuned \cite{Akira2019}. 

Several experimental in vitro setups have been proposed to study the response of active cytoskeletal networks to externally applied stimulation, ranging from patterned surfaces and changing confinement geometry to embedded beads manipulated by optical and magnetic tweezers and interactions at water/oil interface \cite{tanimoto2018physical,ross2019controlling, laan2012cortical,Opathalage2019,Guillamat2017}. 
In many of these setups active networks of polymerised microtubules and kinesin-1 motor proteins have been used as model system for investigation on cytoskeletal structures and active matter \cite{DogicReview2017}. 
In these networks, microtubules are randomly oriented in the bulk and brought out of equilibrium by kinesin motors that bind to the filaments and move along them, exerting active stress during the stepping process, driven by the continuous supply of adenosine triphosphate (ATP) in solution. The activity of the motors is supported by arranging the microtubules into bundles through the presence of depleting agents, such as polyethylene glycol (PEG), which causes an effective attractive interaction between the filaments by entropic force \cite{asakura1958interaction,needleman2004synchrotron}. Under this arrangement the motors 
generate contractile and extensile forces that lead to spatial displacement of the microtubules and bending of the bundles on a larger length scale as well as to emergent behaviour characterised by contraction, formation of microtubule asters, vortices or nematic structures in 2D and 3D \cite{surrey2001physical,ndlec1997self, sanchez2012spontaneous, strbing2020wrinkling}, resembling biological processes in cells.

In this study, we experimentally investigate such active microtubule-kinesin networks and their pattern formation under external mechanical stimulation. For this purpose, we confine the active networks in an evaporating sessile droplet, which provides an aqueous compartment that establishes controlled environmental conditions for our biological network, similar to its natural context. Importantly, the droplet enables the mechanical stimulation of the network by a combination of forces of different nature: The free surface not only confines the network due to its capillary action, but also drives shear flows in the droplet. These flows exert stresses on the entire active microtubule network and couple with its intrinsic activity. This configuration resembles the external forces that cytoskeleton is occasionally exposed to.
In particular, evaporative losses are compensated by outward capillary flows, but preferential evaporation leads to compositional gradients that, in addition, drive inward Marangoni flows.
Aggregation of colloids into ordered or disordered structures under such flows have been the topics of many studies \cite{deegan1997capillary,pearson_1958, berg_boudart_acrivos_1966,Cloot1990,marin2011order,marin2012building,marin2019solutal}. 
Here, such a flow field exerts (shear) stresses on an active, biological system, namely the microtubule-motor protein network, and competes with the active stresses in the network. This leads to the emergence of pattern as they occur in purely biological systems under stress.

\section*{Results and Discussion}
\subsection*{Dynamics of active microtubule networks in hydrodynamic shear flow}
\begin{figure*}[ht!]
	\begin{center}
		\includegraphics[width=\textwidth]{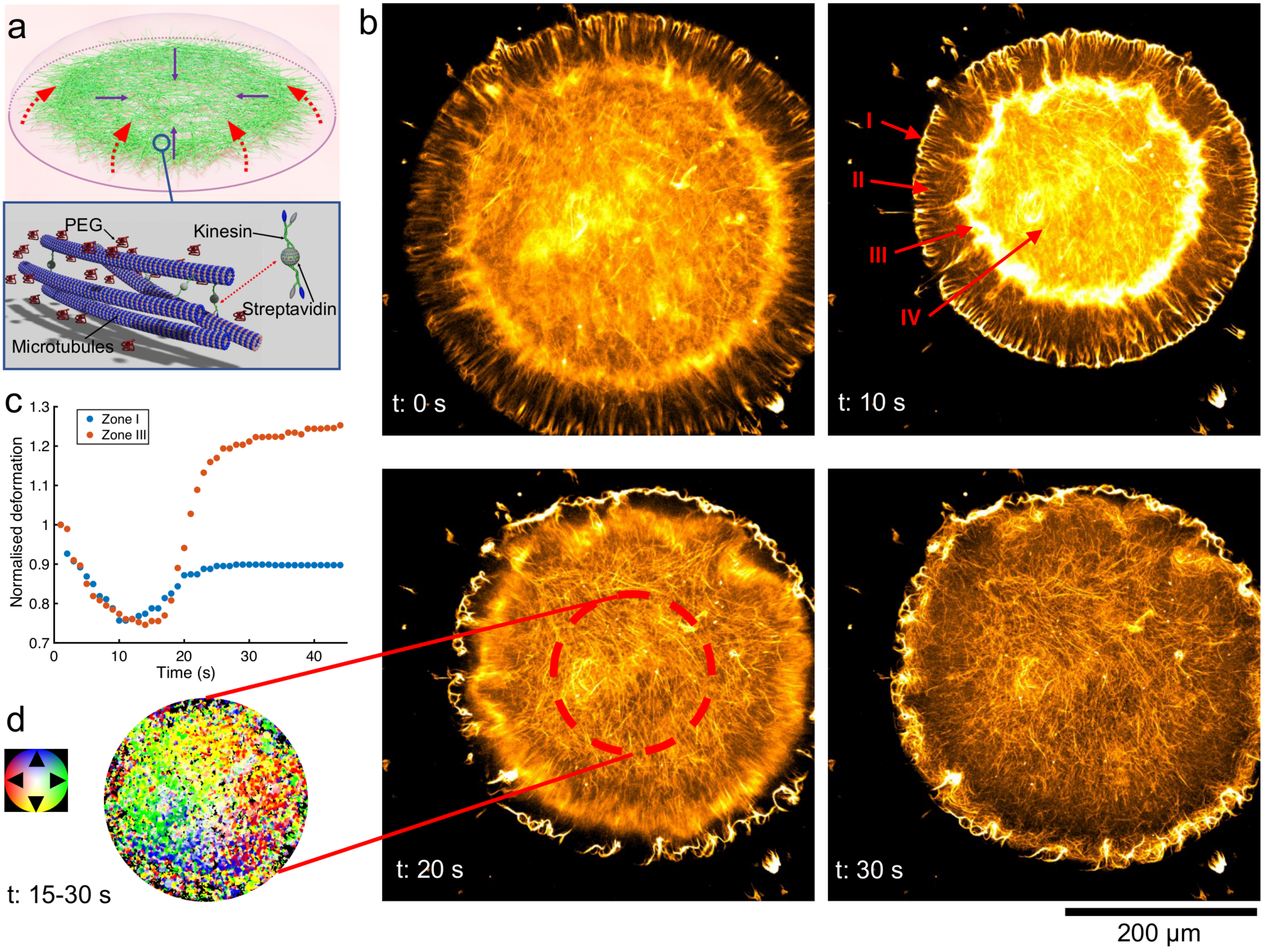}
	\end{center}
	\caption{Dynamics of active microtubule network inside an evaporating droplet. a) Schematic representation of the experimental setup. Microtubule active network embedded into a droplet of a mixture of buffer and PEG on a PLL-g-PEG functionalized substrate. The inset shows the biological building blocks constituting the network, namely microtubules and kinesin-streptavidin motor clusters, arranged into bundles due to depletion force. b) Micrographs showing the emergent behaviour of the self-organizing active network during the droplet evaporation over a time of 30 s. The pattern can be described by four zones marked as I, II, III and IV. c) Dynamics of the outer and inner boundary of the droplet during evaporation corresponding to zone I and zone III, respectively. d) Results of optical flow analysis of the active microtubule network in zone IV during the relaxation phase (t = 15--30 s). It shows inward activity of the microtubule network at the centre of the droplet.
		\label{fig:fig1}}
\end{figure*}

A droplet, extracted from an active mixture mainly composed of stabilized microtubules, motor proteins kinesin-1, PEG and ATP (prepared as explained previously \cite{strbing2020wrinkling}) was deposited onto a PLL-g-PEG functionalised glass slide (see Material and Methods for details). Such a system is shown in the schematic representation in Figure \ref{fig:fig1}.a. We could observe an interconnected network of rod-like shaped microtubules, which self-assembled in bundles due to the addition of PEG. The average length of the filaments was around 15 $\mu$m, but the bundles that they formed exceed this length by an order of magnitude. The microtubules had fluorescent label to allow their visualisation. The addition of motor proteins kinesin-1 allowed the cross-linking of the microtubules within the bundles and of the bundles with each other (see inset in Figure \ref{fig:fig1}.a ), arranging them into a network. The motor proteins move along the microtubules, causing active, i.e., non-equilibrium stresses in the network. Hereafter we refer to this system as an active droplet. 

We observed the continuous evolution of this active network by recording its fluorescence signal while the droplet evaporated into the ambient atmosphere. In this configuration, the microtubule network was subject not only to the forces exerted by the motor proteins but also to confinement forces between substrate and free surface of the droplet, as well as drag forces generated by the flow in the droplet. This combination led to a unique pattern of network density and orientation of microtubule bundles. The network was visualised by fluorescence microscopy, shown in Figure \ref{fig:fig1}.b for several instants during the evolution of the active droplet. Droplet evaporation began immediately after depositing the mixture onto the glass slide, and visualisation started after a few seconds, corresponding to t~=~0~s in the figure.
The pattern generated by the network in the droplet can be divided into four distinct radial sectors according to the fluorescence signal in Figure \ref{fig:fig1}.b (labelled I-IV for t = 10 s). At t~=~0~s and t~=~10~s, all four zones  are simultaneously visible. The outermost zone (I), where the fluorescence signal abruptly vanishes, represents the contact line of the droplet along which filaments likely accumulate tangentially, visualized by a narrow bright ring at the very edge (more evident at t = 10 s).
Notably, no filament deposit was left behind during the receding contact line motion, likely due to the combination of Marangoni circulation~\cite{Leenaars:Langmuir1990} and the PLL-g-PEG functionalised glass substrate, which reduced the unspecific protein adsorption on the surface.
Next to the contact line, we find an annular region (II) in which radially oriented microtubule bundles are rather uniformly distributed around the droplet.
Toward the drop center follows a ring of high fluorescence intensity without visible bundle orientation (III).
In the fourth, innermost sector (IV, covering the center of the droplet), isotropic randomly oriented microtubule bundles form a dense network.
The emergence of these distinct zones is caused by the combination of active forces at the molecular scale, effectively contracting the network, and drag forces from the surrounding flow field that stretch and shear the network.
Qualitatively, the dynamics of the system is characterized by an initial contraction, followed by a spreading motion of the droplet footprint. Simultaneously, the network moves relative to the footprint, in the same direction, effectively amplifying its motion (cf. Fig.~\ref{fig:fig1}.c, comparing the contact line (zone I) and high fluorescence intensity ring (zone III)). 
Movie S1 (Supplementary Information) comprehensively visualizes this dynamics.
Figure \ref{fig:fig1}.c shows the pattern dynamics over the evaporation time. During the first $\sim 10-15$~s after recording was started (images at $t=0$~s and $t=10$~s in Fig.~\ref{fig:fig1}~b), the droplet footprint diameter shrunk by about $\sim 24$\%. 
Zones III and IV, i.e., the isotropic network in the droplet center and the high fluorescence intensity ring around it, showed a similar contraction (Figure \ref{fig:fig1}.b, t = 10 s, Figure \ref{fig:fig1}.c). 
Afterwards, until $t\sim 30$~s, the droplet spread again to a larger footprint diameter up to $\sim 90$\% of its initial size (Figure \ref{fig:fig1}.c). The annulus of radially oriented microtubule bundles (zone II) gradually disappeared in this phase, the bundles from this region have been advected into the contact line (zone I), which became more irregular.
The high fluorescence intensity ring (zone III), now also somewhat more irregular, expanded faster to catch up with the contact line, exceeding its initial size (Figure \ref{fig:fig1}.c). During these later stages, the isotropic network (zone IV) expanded through the entire droplet (Figure \ref{fig:fig1}.b at t = 30 s). The faster expansion of the high fluorescence intensity ring (zone III) compared to the contact line (zone I) was evident also by considering their spreading velocity (See Figure S1 in SI).
However, a weak contraction was still visible near the center of the droplet.
We quantified this contraction by estimating the optical flow between images (Figure \ref{fig:fig1}.d), which reveals the inward motion of fluorescent filament bundles in the central region, even before the active droplet stopped spreading. 

\begin{figure*}[ht!]
	\centering
	\includegraphics[width=0.9\textwidth]{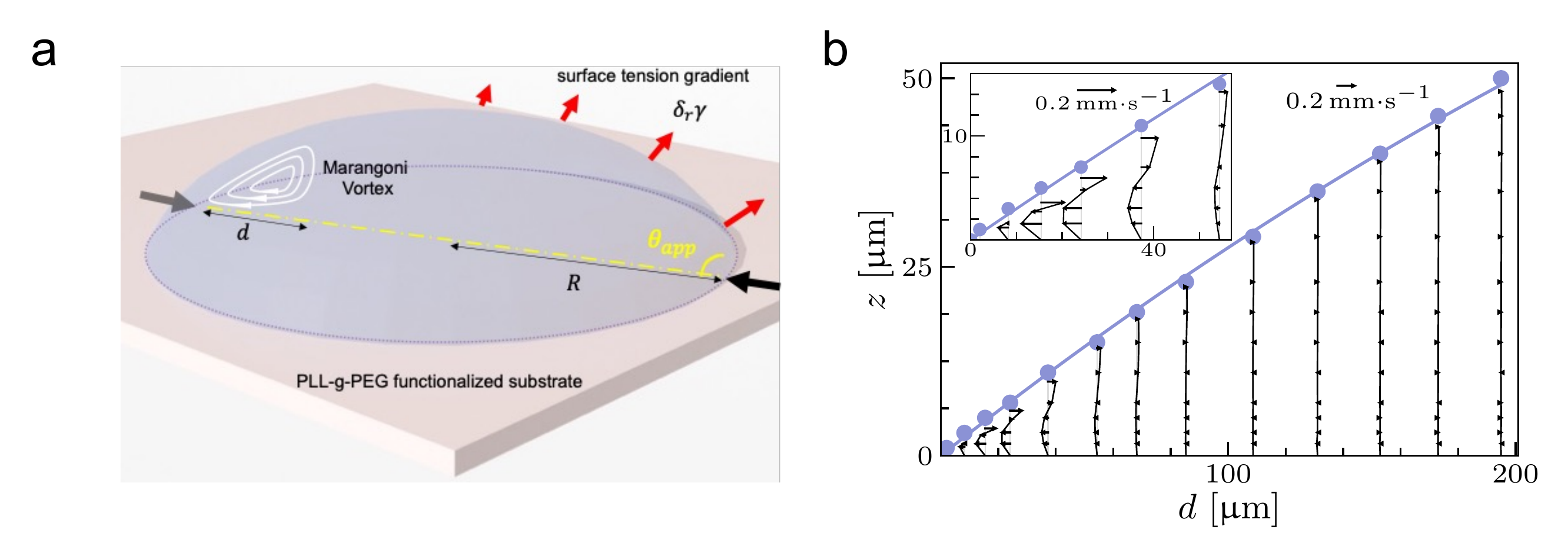}
	\caption{a) Schematic representation of an evaporating droplet and PEG-induced Marangoni flow on a PLL-g-PEG functionalised surface. b) Measurements of the internal flow velocity of the droplet obtained with high resolution micro particle image velocimetry. Cross-sectional view of the droplet composed of M2B and PEG with molecular weight of 6 kDa: free interface (blue circles and fitting line), radial velocity (black arrows) and velocity profiles (black lines). Inset: close up to the Marangoni flow, locally formed near the contact line.}
	\label{fig:fig3}
\end{figure*}

We explain the emergence of the observed pattern and dynamics by the interplay of the active stresses exerted by the motor proteins on the microtubules and the drag forces from the flow field. 
Namely, the PEG in the solution acts as a depletion agent to bundle the microtubules, enhancing the binding of kinesin motors and thus the force generation by their active motion along the microtubules.
The interconnected microtubule network experiences contraction and extension depending on the polarity of the filaments in the bundles.
However, PEG also acts as a surfactant to the aqueous solution. Enriched near the contact line due to water evaporation, PEG induces Marangoni flows in the droplet (Figure~\ref{fig:fig3}). The shear stresses from these flows couple to the active stress, acting against the forces from the molecular motors, which leads to the arrangement into the different zones.
Specifically, evaporation is strongest near the edge of the droplet, where, in addition, the droplet also becomes very shallow. Thus, PEG enrichment is strongest near the contact line, where surface tension becomes lowest. A surface tension gradient develops (Figure \ref{fig:fig3}.a), driving a Marangoni flow close to the free surface which is directed toward the center of the droplet. This flow is compensated by an outward capillary flow, driven by evaporation and minute changes in surface curvature, close to the substrate, similar to what is known from Marangoni contraction~\cite{cira2015vapour,ramirez2022taylor,baumgartner2022marangoni}.
We verified such PEG-induced Marangoni flows inside the droplet by analysing the evaporation of a droplet composed of buffer (see Materials and Methods section) and PEG, but without the active network (Figure \ref{fig:fig3}.b and Figure S2).
Z-resolved micro-PIV measurements indeed revealed a Marangoni vortex which was contained within $\sim$ 50 $\mu$m from the contact line.
The accumulation of PEG molecules at the free surface of the droplet was further increased by the salty buffer (M2B containing mostly MgCl$_{2}$), known from polymer and salt-induced condensation \cite{yoshikawa2010compaction,xu2012polymer,cheng2015polyethylene}.
We confirmed this PEG-salt interaction by measuring the surface tension at different PEG and salt concentrations (See Table S1 in Supporting information). 
The consequence of this Marangoni vortex was a zone with strong shear forces which coincides with the region where radially oriented bundles (zone II) were observed in the active droplet.
Apparently, the coupling of strong shear to the active forces generated by the molecular motors was responsible for the alignment of the active bundles: This alignment was not observed in absence of the motors, as we verified in a control experiment (see below).
Next to the vortex, toward the drop center, the flow stagnates, leading to an accumulation of material at the boundary with the active network, which pushed against this accumulation, forming the bright ring (zone III). In the central region (zone IV), flows are weak and the network activity dominated.
Interestingly, we did not observe that the active network in the central region collapsed into an aster-like structure as it has been observed in isotropic networks of microtubules and kinesin motors in closed chambers~\cite{nedelec1997self}.
Rather, it continued to fill zone IV and likely resisted the radial shear stress of the Marangoni flow by exerting a force outwards. 
We hypothesise this is due to the dual nature of the forces that kinesin clusters generate within the microtubule bundles. Specifically, depending on the polarity of the filaments that the motor proteins cross-link, the kinesin can exert also extensile and not exclusively contractile forces \cite{sanchez2012spontaneous,strbing2020wrinkling},
which seems to be linked to the resistance to the shear flows. This phenomenon resembles the behaviour of the microtubule network inside the cell, where it has been shown that it can withstand large-scale compressive forces \cite{Brangwynne}, providing cellular mechanical stability and balancing the contractile stress \cite{wang2002}.

The pattern persisted as long as the Marangoni flow was strong enough. As evaporation continued to remove water from the solution, the PEG concentration increased until the surface was saturated and the Marangoni flow weakened.
It is known that the intensity of Marangoni flows exhibit a maximum with respect to PEG concentration~\cite{cheng2015polyethylene}. 
At this point the hydrodynamic aggregation caused by the drag forces acting on the active network also disappeared while the capillary flow prevailed, leading to the outward movement observed inside the droplet. We found an analogous non-monotonic trend of the contact angle of evaporating droplets that contained only PEG. Coupled to the strength of the Marangoni flow, the contact angle reached a maximum and then decreased (See Figure S3 in SI). 
Once the Marangoni flow decreased, the network spreads as observed during the experiments (Fig. \ref{fig:fig1}.b).
The residual contraction activity of the network in the central region of the droplet is evidently due to the action of the motors, which was less affected by the radial shear stress (Figure \ref{fig:fig3}.b). Thus, the network continued to contract toward the center of the droplet. However, the motor activity appeared weaker compared to the active stress that we observed in previous studies \cite{nasirimarekani2021tuning,strbing2020wrinkling} for kinesin-1 working in ATP saturation regime. We conjecture that the reduced activity in the center of the droplet and completely disrupted elsewhere was the result of the altered concentration of PEG and salt due to the water loss by evaporation. Specifically, increasing the salt concentration affects the motor protein activity in several ways: high concentration of salt results in detachment of kinesin from microtubules \cite{thorn2000engineering} and suppresses the ATP hydrolysis \cite{gilbert1995pathway}, both impeding the activity of motor proteins. Additionally, increasing the PEG concentration in the solution raises the depletion force. This results in higher attractive forces between the microtubules, which hinder the free ``walking'' motion of the motors. Both, the flows in the late stage and the reduced network activity, led the system to relax as we observed during our experiments.

\subsection*{The role of PEG in the network contraction} 
\begin{figure*}[h!]
	\centering
	\includegraphics[width=0.9\textwidth]{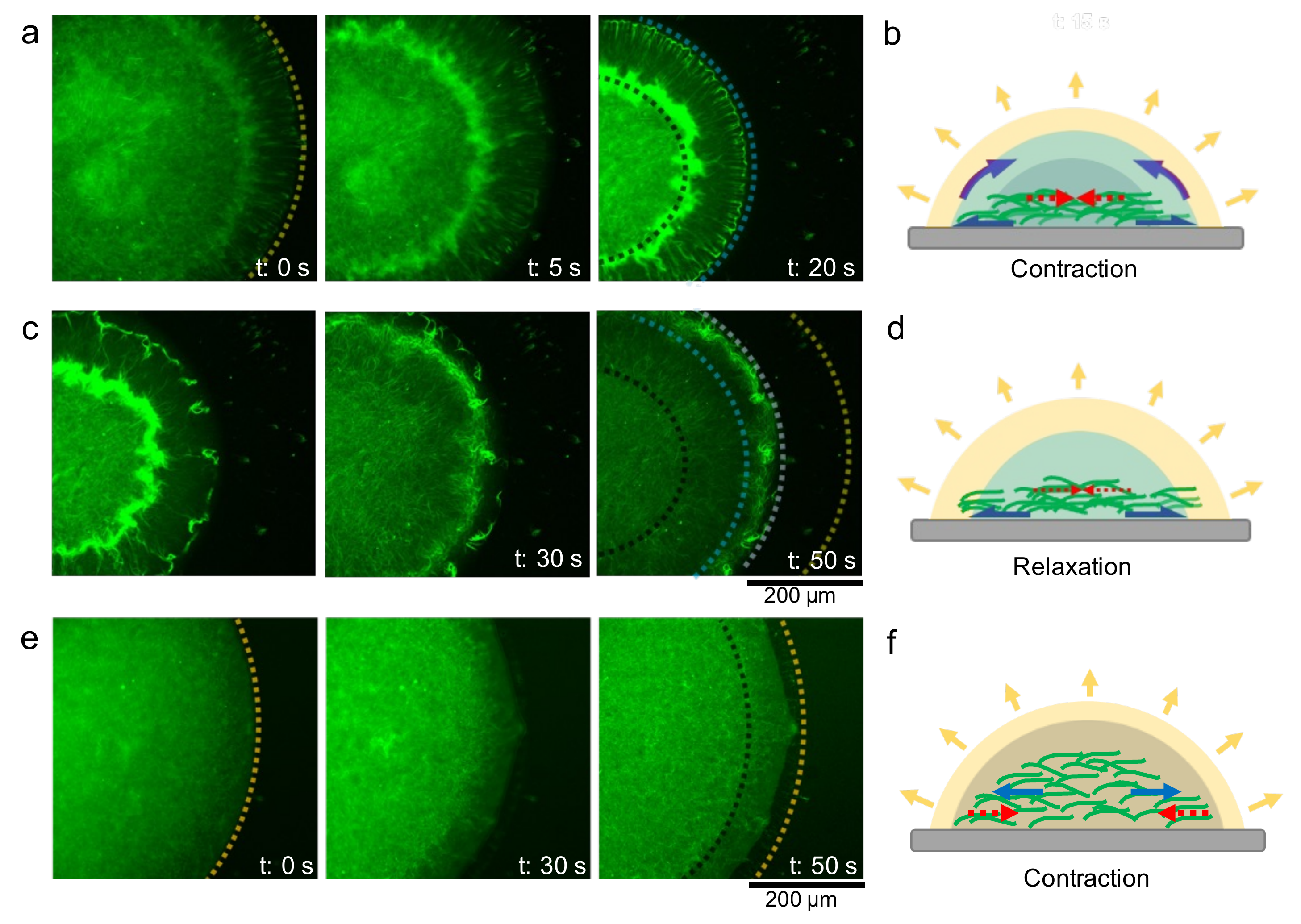}
	
	\caption{Comparison of the active microtubule network in the presence (a-d) and absence (e, f) of PEG. a, b) Micrographs and schematics of the droplet containing PEG with molecular weight of 20 kDa, for time interval of t = 0 -- 20 seconds, showing an inward contraction. Yellow dash line refers to the initial boundary of the evaporating droplet, blue and black dash lines show final contracted outer contact line and inner ring of densely contracted microtubule bundles, respectively. c,d) Micrographs and schematics of the droplet containing PEG under relaxation, for time interval of t = 30 -- 50 seconds. e,f) Micrographs and schematics of active network in a droplet without PEG, showing a continuous inward moving activity of the microtubule network at the contact line of the droplet.}
	\label{fig:fig2}
\end{figure*}

As a depletant, PEG plays a pivotal role in the dynamics of these active systems, which we tested by repeating the experiments without PEG in the active mixture. In Figure \ref{fig:fig2}, we compare the results of this setup (panels e, f) with the dynamics of the fully active droplet as described above (panels a-d).
In the absence of PEG, the active network inside the evaporating droplet showed a different behavior already at the beginning of the experiments (Figure \ref{fig:fig2}.a vs Figure \ref{fig:fig2}.e and Movie S2 vs Movie S3). Without PEG we did not observe the characteristic pattern of four different arrangements, nor its evolution over time, as described above and shown again in Figure \ref{fig:fig2}.a and c.
Without PEG we observed a continuous movement of microtubules toward the center of the droplet and a slight receding contact line motion (Figure \ref{fig:fig2}.e and Movie S3). However, neither the contact line motion nor the aggregation of the active network were as pronounced as in the droplet with PEG (Figure \ref{fig:fig2}.a). The annulus with radially organised bundles and the high fluorescence intensity ring were completely absent.
With the schematic representations in Figure \ref{fig:fig2}.b,d and f we motivate the force fields inducing the contraction/aggregation in the early phase (a) and the relaxation afterwards (b) for the case including PEG, and the single phase observed without PEG (f).
Without PEG, no Marangoni flows were generated inside the droplet. The contraction was exclusively generated by the motor proteins, competing with the capillary flow due to evaporation (Figure \ref{fig:fig2}.f). 
But also the absence of PEG as a depletion agent is important. Since the microtubule solution did not undergo phase separation into bundles, the network was not as interconnected as in the presence of PEG.
This plays a decisive role for the emergent behaviour in the active droplet.
If the network consists of single filaments, cross-linked by the motor proteins, but without depletion that could aggregate them, the force transmission is not as efficient as in the case with PEG. We conclude that the PEG is one of the main actors in the system that is required to build active droplets. It establishes the optimal configuration of the active microtubule network that allows the motors to efficiently generate stress. Additionally, it generates Marangoni flows that cause aggregation patterns of the biological network on larger length scales. 
\subsection*{The role of motor proteins in pattern formation}
\begin{figure*}[h!]
	\centering
	\includegraphics[width=\textwidth]{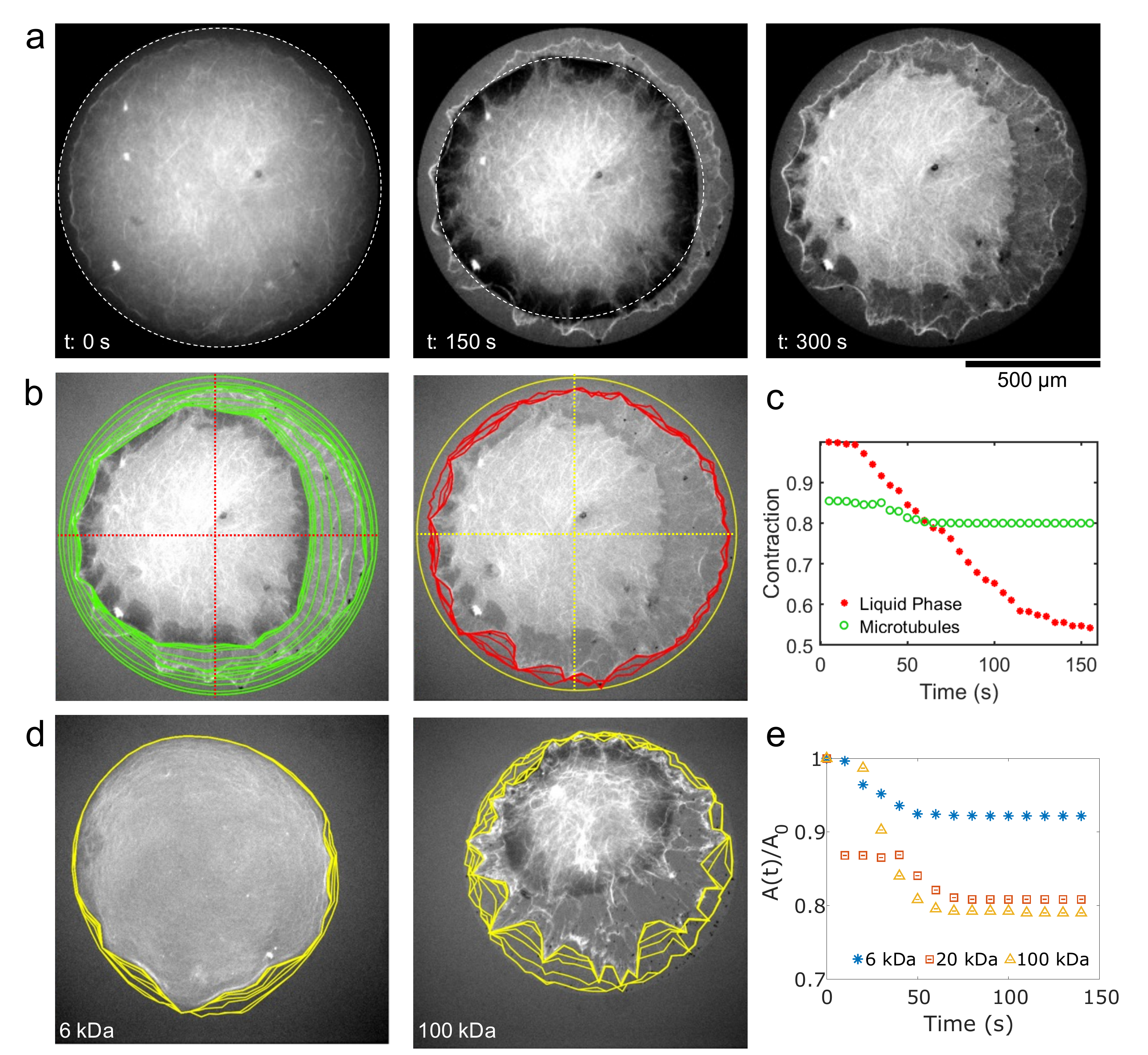}
	\caption{Aggregation behaviour of passive microtubule network inside an evaporating droplet (in the absence of motor proteins). a) Micrographs of the evaporating droplet containing PEG with 20 kDa molecular weight, over the course of 5 minutes. White dash line represents the contact line of the droplet as boundary for measuring its surface area over time, $A(t)$. b-c) PIV results and quantitative analysis of the contraction of internal dense region and entire network over time. d) Contraction of droplet with passive microtubule network and PEG with 6 kDa (left) and PEG 100 kDa (right) molecular weight. e) Graph of the aggregation of the evaporating droplet over time, defined as $A(t)/A_{0}$, for mixtures with PEG of different molecular weight (6, 20, and 100 kDa).}
	\label{fig:fig4}
\end{figure*}
While PEG plays a crucial role in building both the optimal conditions for active stress transmission within the system and the fluid force field inside the evaporating droplet, the major actor in the out-of-equilibrium activity of the biological network is the action of motor proteins kinesin-1.
In order to show its contribution in the coupling between force exerted by the molecular motors and by the shear flow inside the evaporating droplet, we tested the system including the PEG in the sample mixture but without adding motor proteins, i.e., a passive network of bundled microtubules. Interestingly, this led to quite distinct dynamics both in space and time during the droplet evaporation. At the beginning of the experiment (Figure \ref{fig:fig4}.a, t = 0 s) we observed a dispersion of randomly distributed microtubule bundles inside the droplet, which started to evaporate and shrunk over time (Figure \ref{fig:fig4}.a, the dashed line indicates the moving contact line of the droplet). 
Over the evaporation time a very peculiar configuration was observed. The network and the droplet did not show any synergy as in the active case. The droplet footprint shrunk through a stick-slip mechanism, with an inward movement relative to the microtubule network. The latter partly adhered mechanically to the substrate, while the other part was aggregated by the shrinking droplet compartment (visible in Figure \ref{fig:fig4}.a -- t = 150 s -- as the high intensity fluorescence area). Note that the substrate was functionalised identically as in the active cases described above.
We conjecture that the cross-linking capability and activity of the motor proteins in the active case is required to overcome the adhesion of the network to the substrate. Then, the Marangoni circulation near the contact line additionally prevents deposition and pinning. However, without the kinesin action the Marangoni circulation is not sufficient to compensate for the mechanical interaction of the network with the substrate.
Thus, only the collaboration between kinesin forces and a surrounding flow field can generate the patterns observed in the active case. 
The droplet continued to contract further and dried out completely. Finally, a fluorescent network with a higher filament density in the central region and a less dense region around it (Figure \ref{fig:fig4}.a, t = 300 s) was left behind. The high density region and the entire network system were tracked over time and we quantified their corresponding contraction (Figure \ref{fig:fig4}.b,c). It is clearly visible that in the passive case the network was not pushed symmetrically in the middle of the droplet (Figure \ref{fig:fig4}.b). The internal network region aggregated by more than 40\%, while the entire passive network reached a plateau of 20\% quickly at the beginning of the experiment.
Overall, the resulting process was slower compared to the active droplet, indicating that the motor proteins define both the length and the time scale of the system dynamics.
We also tested the influence of the distribution and size of microtubule bundles on the aggregation behaviour of the passive network by using PEG with different molecular weight, e.g. 6 and 100 kDa. We observed that shorter PEG chain length induced thinner bundles that likely resulted in a less meshed network. This is clearly visible in Figure \ref{fig:fig4}.d by comparing the network generated by PEG with molecular weight of 6 kDa and 100 kDa with that generated by PEG with molecular weight of 20 kDa (Figure \ref{fig:fig4}.b).
We measured the aggregation ratio (defined as $A(t)/A_0$, with $A(t)$ the network surface aggregated over time and $A_0$ that at t = 0 s) for passive networks obtained by using PEG with the three different molecular weights. Figure \ref{fig:fig4}.e shows a linear relationship of the aggregation ratio with the PEG molecular weight, with the 6 kDa PEG network achieving an aggregation of 10\% and the 20 kDa and 100 kDa PEG networks reaching a stronger aggregation around 20\%. This is due to a stronger Marangoni flow inside the droplet contact line at higher PEG molecular weight, which was confirmed by PIV measurements (See Supplementary Methods and Figure S2 in SI).
However, in all three cases an asymmetric and seemingly random aggregation pattern was observed, unlike in the active droplet experiments, which showed symmetric self-assembly and a ring-like zones of distinct patterns. 
This asymmetric positioning of the network inside the droplet could be due to the non-homogeneous distribution of microtubule bundles inside the network. In other words, the missing kinesin activity inside the network did not allow both the cross-linking of the system and the isotropic transmission of the contraction force that re-arrange the microtubule network under the mechanical stimulation of the external radial stress.
We confirmed that this behaviour was exclusively due to the presence of the microtubule in the solution by repeating the experiment and excluding the biopolymer network. In this case the droplet contracted with a circular shape towards the center of the droplet. We also repeated the experiment and included latex particles. In this case we observed a behaviour similar to the droplet containing microtubules (See Figure S4 in SI).

\subsection*{The role of the functionalised substrate}
\begin{figure*}[h!]
	\centering
	\includegraphics[width=0.8\textwidth]{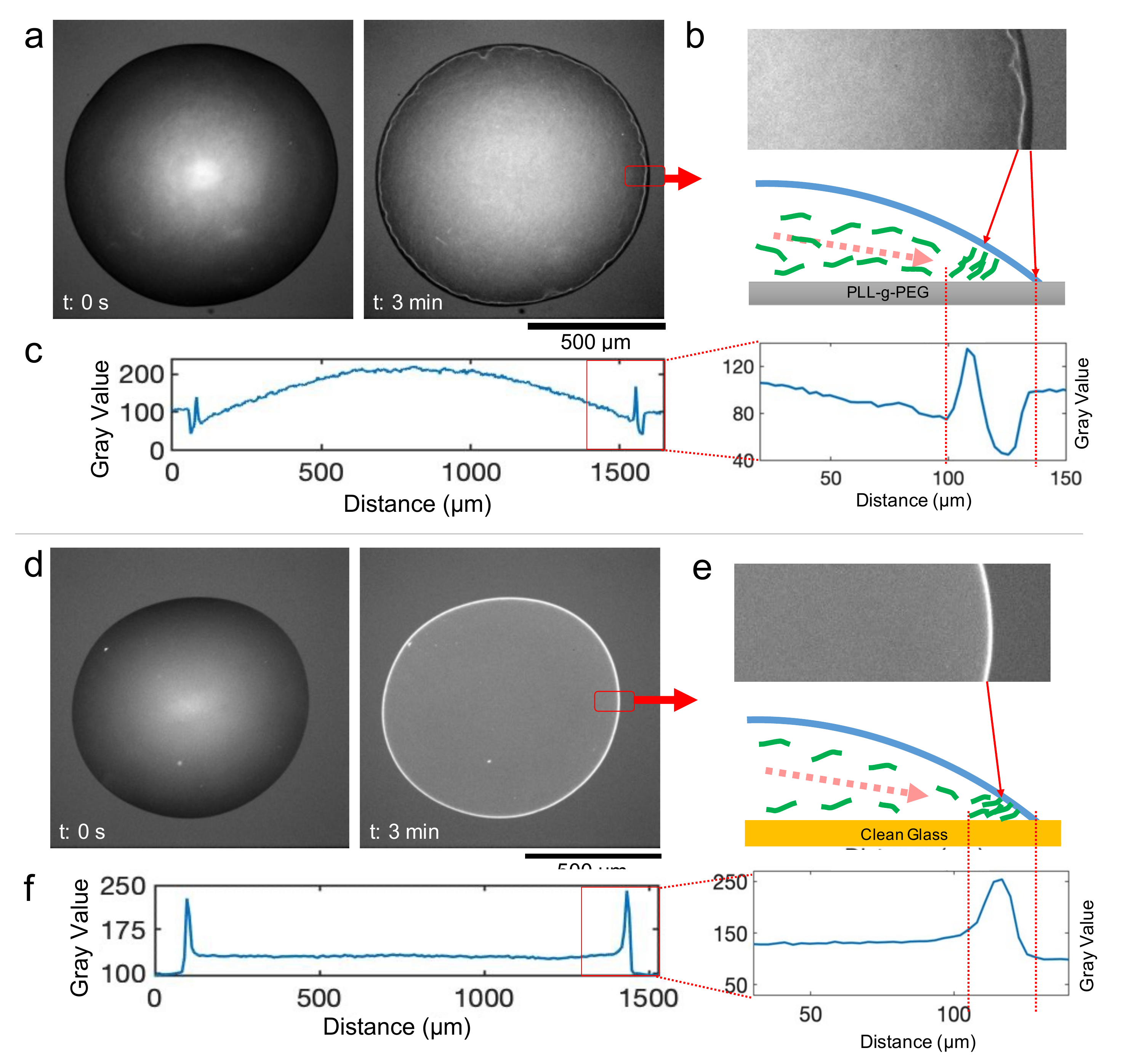}
	
	\caption{Comparison of the classical coffee stain formation by passive microtubules in droplets without PEG, on functionalized and non-functionalized glass slides. a) Micrographs of the evaporating droplet on PLL-g-PEG functionalized glass surface. b) Schematics showing the coffee stain of microtubules on functionalized glass slide. Gray value measurements depict a gap between the coffee stain and the contact line. c) Micrographs of the evaporating droplet on non-functionalized glass surface. d) Schematics showing a profile view of the coffee stain of microtubule on non-functionalized glass slide. Gray value measurements indicate the accumulation of the filaments at the exact contact line.}
	\label{fig:fig5}
\end{figure*}

For all the experiments presented in this study we used glass substrates that we
coated with PLL-g-PEG. The functionalization reduced the protein adsorption on the substrate, e.g., adhesion of microtubule filaments and motor proteins. Indeed, during evaporation of the active droplets no deposition could be observed at the contact line. We were interested in understanding how reducing the interaction of the proteins with the substrate might affect the dynamics of the system and its behaviour at the contact line of the droplet (Figure \ref{fig:fig5}). For this purpose the evaporation of a passive droplet without PEG was analysed on functionalized and non-functionalized (clean glass, see Materials Methods for more details) substrates. We excluded PEG to avoid Marangoni flow and studied the interaction of microtubules under the capillary flow. 
We compared the initial and final stage of evaporating droplets for the two surfaces, i.e., functionalised (Figure \ref{fig:fig5}.a) and non-functionalised substrate (Figure \ref{fig:fig5}.d). At the onset of the experiments the two droplets looked similar (Figure \ref{fig:fig5}.a and d, t = 0 s). After 3 min of evaporation, the different effects of the capillary flow on the passive microtubule network for the two surfaces was visualised by the fluorescence light intensity of deposition at the droplet contact line.
On functionalized glass, we observed filament deposition that did not overlap with the droplet contact line (Figure \ref{fig:fig5}.a t = 3 min). Specifically, a gap in the fluorescence intensity could be seen between the network and the contact line (Figure \ref{fig:fig5}.b).
We hypothesise that the gap was caused by the improved spreading of the droplet and the repulsion of the biopolymers from the functionalized surface. This hypothesis was tested by repeating the experiments with spherical particles (950 nm in diameter) instead of proteins (microtubules). An evaporating droplet containing spherical particles, which are much bigger than microtubules and insensitive to the repulsive effect of the functionalization, accumulated exactly at the contact line, in either case (See Figure S5 in SI). 
When we used a non-funtionalized substrate for the experiment, a clear ring immediately at the contact line of the droplet could be observed (Figure \ref{fig:fig5}.d). There, the fluorescence intensity analysis showed no gap between microtubule and the edge of the droplet (Figure \ref{fig:fig5}.e), reminiscent of the coffee-ring effect. Furthermore, a uniform fluorescence intensity over the entire droplet footprint emerged after evaporation, which corresponds to a homogeneous deposition of unspecifically adhered microtubule on the internal surface (Figure \ref{fig:fig5}.f).
These different outcomes show that also the interaction between the microtubules and the substrate plays an important role in the emergent phenomena of the entire system.

\section*{Conclusion}

We studied the dynamics of an active microtubule network inside an evaporating droplet on a solid surface at room temperature. Under these conditions the coupling of the intrinsic activity of the biological network with the flow field generated inside the droplet could be analysed. PEG, the depletion agent used to bundle the microtubules, induced Marangoni flows that contributed to the contraction of the active network. A distinct pattern emerged as a result of this combination of forces. We have shown by systematic reduction that PEG, kinesin motors, microtubules and surface functionalisation all play distinct and important roles in the pattern formation. Namely, PEG is required to induce the bundling of the microtubules and thereby hone the network architecture for an efficient active force generation. Equally important, PEG induces Marangoni flows that apply radial shear stresses on the actively contracting network, leading to zones of distinct large-scale organization. The kinesin motors are the main contributors for the cross-linking and activity of the network inside the droplet. Indeed, by omitting the motors from the solution we have shown that the network phase separated from the aqueous phase of the droplet, as the  adhesion to the substrate could not be overcome by the flow field. Finally, we have shown that the interaction between biopolymers and substrate also contribute to the determination of the collective behaviour.
The evaporating droplet offered an alternative and innovative environment to stimulate active biological networks and analyse their response to applied stresses. With a non-invasive and simple methodology, we could generate behaviour that are typical of microtubule cytoskeleton in the intracellular space, like opposing resistance to large-scale contraction \cite{Wang2001}, actively reacting to the membrane deformation \cite{Brangwynne} and balance contractile stress \cite{wang2002}. Furthermore, these biopolymer networks hold great potential to revolutionize the field of biomaterials in the future. Here we show that they exhibit interesting features that can be easily tuned and controlled.

\section{Materials and Methods}
	
	We briefly describe the experimental procedures and the visualisation and analysis tools used in this study. More detailed descriptions can be found in the Supplementary Information.
	\subsection*{Polymerization of microtubules}
	Microtubules were polymerized from 2.7 mg/ml HiLyte labeled
	porcine brain tubulin (Cytoskeleton, Inc., U.S.A.) in M2B with 5 mM MgCl$_{2}$, 1 mM GTP, and 5\% DMSO
	at 37$^{\circ}$C for 30 min. The microtubules were stabilized with 7 $\mu$M taxol and mixed with 0.5 mg/ml glucose, 0.65 mM dithiothreitol (DTT), 0.2 mg/ml glucose oxidase (Sigma G2133), 0.05 mg/ml catalase (Sigma
	C40) and 2.4 mM Trolox (Sigma 238813) to avoid photobleaching. Depending of the specific case, 1\% PEG with molecular weight 6, 20 and 100 kDa was added. For the experiments on active droplets kinesin 401 was added. The plasmid that codes biotin-labeled kinesin 401 (K401) was a gift from Jeff Gelles (pWC2 - Addgene plasmid \# 15960; http://n2t.net/addgene:15960; RRID Addgene 15960)\cite{Subramanian445}. Kinesin 401 was purified as previously published\cite{Gilbert,Young} and the kinesin-streptavidin complexes were prepared by mixing 0.2 mg/mL kinesin 401, 0.9 mM DTT, and 0.1 mg/ml streptavidin (Invitrogen, S-888) dissolved in M2B and
	incubated on ice for 15 min. 4 $\mu$l of this mixture was mixed with ATP at a final concentration of 1 mM, 1.7 $\mu$l of pyruvate kinase/lactic dehydrogenase (PK/LDH, Sigma, P-0294), 32 mM phosphoenol pyruvate (PEP, VWR AAB20358-06) to form a solution of active clusters that was added to the microtubules solution described above.  The microtubules and the active clusters solution were mixed 15 min before the onset of the experiments.
	
	\subsection*{Functionalized glass surface preparation}
	Microscope coverslips were cleaned by washing with 100\% ethanol and rinsing in deionized water. They were further sonicated in acetone for 30 min and incubated in ethanol for 10 min at room temperature. This was followed by incubation in a 2\% Hellmanex III solution (Hellma Analytics) for 2 h, extensive washing in deionized water, and drying with a filtered airflow.
	When the functionalisation of the substrate was required, the cleaned coverslips were immediately activated in oxygen plasma (FEMTO, Diener Electronics, Germany) for 30 s at 0.5 mbar and subsequently incubated in 0.1 mg/mL poly(l-lysine)-graft-poly(ethylene glycol) (PLL-g-PEG) (SuSoS AG, Switzerland) in 10 mM HEPES, pH = 7.4, at room temperature for 1 h on parafilm (Pechiney, U.S.A.). Finally, the coverslips were lifted off slowly, and the remaining PLL-g-PEG solution was removed for a complete surface dewetting.
	
	\subsection*{Image acquisition and analysis}
	Image acquisition was performed using an inverted fluorescence microscope Olympus IX-71 with a 4$\times$, 10$\times$ or 20$\times$ objective (Olympus, Japan), depending on the experimental setup. For excitation, a Lumen 200 metal arc lamp (Prior Scientific Instruments, U.S.A.) was applied. The images were recorded with a CCD camera (CoolSnap HQ2, Photometrics). The frames were acquired at 1 Hz. ImageJ software was used for the analysis of the acquired images. 
	
	\subsection*{PIV measurements}
	The internal flows of evaporating droplets, mixtures of M2B buffer and PEG were quantified using micro particle image velocimetry ($\mu$PIV).
	Three different molecular weights of PEG were used (6, 20 and 100 kDa). The experiments were done inside a humidity control chamber (a cubic chamber with 10 cm size), at room temperature. 
	Droplets with initial volumes of 1 $\mu$L were placed on substrates (functionalized glass coverslides $24\times 24$ mm).
	Polystyrene microspheres (Thermo Fisher Scientific F8809, diameter of 200 nm) were used as flow tracers.
	The particles within the drops were observed with an inverted epifluorescence microscope (Nikon Eclipse Ti2) for the PIV measurements. Please refer to the Supplementary Methods for a detailed description. 
	
\begin{acknowledgement}
	The authors acknowledge support from the Max Planck Society. V.N. and I.G. acknowledge the European Union’s Horizon 2020 research and innovation programme under the Marie Skłodowska-Curie grant agreement MAMI No. 766007. I.G. acknowledges support from the Volkswagen Stiftung ("Experiment!”).
\end{acknowledgement}

\vspace{1 cm}
\noindent The authors declare that they have no competing financial interests.




\bibliographystyle{plainnat}
\bibliography{bibliography}

\newpage

	\section{Supplementary Information}

\singlespacing

\section*{Supplementary Methods}
\subsection*{PIV measurements}
The internal flows of evaporating droplets, mixtures of M2B buffer and PEG were quantified using micro particle image velocimetry ($\mu$PIV).
Three different molecular weights of PEG were used (6, 20 and 100 kDa). The experiments were done inside a humidity control chamber (a cubic chamber with 10 cm size), at room temperature. The humidity was set by continuously injecting a predefined mixture of dry nitrogen and nitrogen saturated with water vapor behind gas-permeable membranes at the side-walls of the chamber. Droplets with initial volumes of 1 $\mu$L were placed on substrates (functionalized glass coverslides, $24\times 24$ mm).
Polystyrene microspheres (Thermo Fisher Scientific F8809, diameter of 200 nm) were used as flow tracers, with a mass fraction of $7.8 \times {10^{-5}}$ of the particle stock solution in the mixtures. The particle concentration was 
$ 1.7 \times {10^{10}}$ particles/mL. 
The particles within the drops were observed with an inverted epifluorescence microscope (Nikon Eclipse Ti2) for the PIV measurements. The microscope was equipped with a water immersion objective (Nikon CFI APO LWD 20-0.95 WI), for diffraction-limited imaging, with a numerical aperture of 0.95. Images of the particles were captured with a high-speed camera (Phantom VEO 4K 990L, imaging speed at 900 fps), quickly switching between planes parallel to the substrate by automating the focus system of the microscope. Once the drop was deposited on the substrate, the surface of the substrate was visible under the microscope by focusing to the particles located at the contact line. After finding this reference plane the recording starts. Thus, the time between the deposition of the droplet and the start of the $\mu$PIV recording is on the order of 10-20 s. A side-view camera (Point Grey Grasshopper2, imaging speed at 27 fps) mounted on the microscope simultaneously recorded the deposition of the drop, the macroscopic apparent shape of the droplet, and the start of the $\mu$PIV recording. The later was detected when the epifluorescence excitation light was observed, after opening the shutter synchronously, due to scattering.  The side-view camera was equipped with a telecentric macro lens (Thorlabs Bi-Telecentric lens 10$\times$\u8192 , working distance 62.2 mm.

\subsection*{PIV measurements analysis}

The recorded images of evaporating droplet for the PIV measurements were analyzed with an in-house developed Python code to quantify the internal flows. Displacements between consecutive images were evaluated by cross-correlation with correlation-averaging over 150 frames (1). An adaptive interrogation window size method is included in the algorithm. First, single-pixel correlations were calculated for the entire image and all displacements within a predefined search range (2). Instead of correlating intensity values directly, we used the dot product of the gradient (first order differences). Then, the correlations were integrated over interrogation windows of various sizes by convolution with a square kernel of the desired size. 
We used five different interrogation window sizes of 4, 8, 16, 32, and 64 px side-length. 
The final correlation maps were then evaluated by a weighted average between the different window sizes, using the mean square of the intensity gradient values in the interrogation window and a size-dependent bias as weight. 
The method was implemented through the Python API of Tensor Flow. 
From this method, 2D velocity fields were obtained for different $z$-planes parallel to the substrate.
The radial component in each $z$-plane was calculated by azimuthally averaging over $100~\mu$m. The corresponding experimental uncertainties of the radial velocities were estimated using the standard deviation over the weighted interrogation windows. With the radial velocities we reconstructed the velocity profiles for different experimental conditions. Close to the free surface, correlations were picked up preferentially from below the focal plane because no particles were outside the drop. This leads to a shift in the correlation plane relative to the focal plane. Accordingly, we applied a correction to the $z$ location of the velocity signal relative to the distance from the free surface.
To obtain this correction, we estimated the point spread function and the correlation sensitivity as a function of distance to the focal plane, and convolved this sensitivity with a unit-step function in $z$ for the particle density. For these measurements, we modeled the depth of correlation as a Gaussian with $1.5~\mu$ m standard deviation.
\subsection*{Contact angle measurements}
The contact angle measurements of the evaporating droplet (M2B+PEG) was conducted by side view image recordings. A digital camera (27 fps; Point Gray Grasshopper2) equipped with a telecentric lens (1.0$\times$; working distance: 62.2 mm; Thorlabs Bi-Telecentric lens) and a collimated light source were used to record the side-view images. The dynamical apparent contact angle $\theta_{app}$ was obtained from the images as $\theta_{app}\simeq2h_0/R$
, where $h_0$ and $R$ are the maximal height and the foot radius of the droplet, respectively (3). 

\subsection*{Interaction of PEG with salty buffer (M2B) - surface tension measurements}
The surface tension measurements were conducted by the pendant drop method (4). For each solution, the surface tension of 8 drops of 2.5 $\mu$L was measured (in room conditions, T = 20 $^{\circ}$C, RH = 45\%). Ten images were collected for each drop in 1 s of recording time. The surface tensions $\gamma_{LV}$ were calculated as an average of these measurements, with an average error of 0.22 mN/m.

\subsection*{References}
1. CD Meinhart, ST Wereley, JG Santiago, A piv algorithm for estimating time-averaged velocity fields. J. Fluids Eng. 122, 285–289 (2000).\\
2. J Westerweel, P Geelhoed, R Lindken, Single-pixel resolution ensemble correlation for micro-piv applications. Exp. fluids 37, 375–384 (2004).\\
3. O Ram\`irez-Soto, S Karpitschka, Taylor dispersion in thin liquid films of volatile mixtures: A quantitative model for marangoni contraction. Phys. Rev. Fluids 7, L022001 (2022).\\
4. F Hansen, G R\o{}dsrud, Surface tension by pendant drop: I. a fast standard instrument using computer image analysis. J. colloid interface science 141, 1–9 (1991).

\section*{Supplementary Figures}
\subsection*{Measurements of contraction and expansion of the active droplet}

Active evaporating droplet underwent contraction and relaxation by displaying a peculiar pattern described in the main text. The coupling between the intrinsic active stress of the biological network and the shear stress by Marangoni flow generated two distinct area: an annulus with ordered arrangement of microtubule bundles surrounding a circular area in which the biopolymer structures contracted as an isotropic distributed network. The  annulus area edges formed on one side the contact line of the evaporating droplet and on the internal droplet side a high fluorescence intensity  ring (Figure \ref{sup.fig:fig1}.a). By tracking the movement of these two edges (also referredto zone I and zone III), we observed that they moved inwardly at similar velocity during the evaporation time. 
However, the velocity during the relaxation period showed a different trend, with the zone III moving at a greater speed compared to the zone I.

\begin{suppfigure}[ht!]

	\centering
	\includegraphics[width=0.9\textwidth]{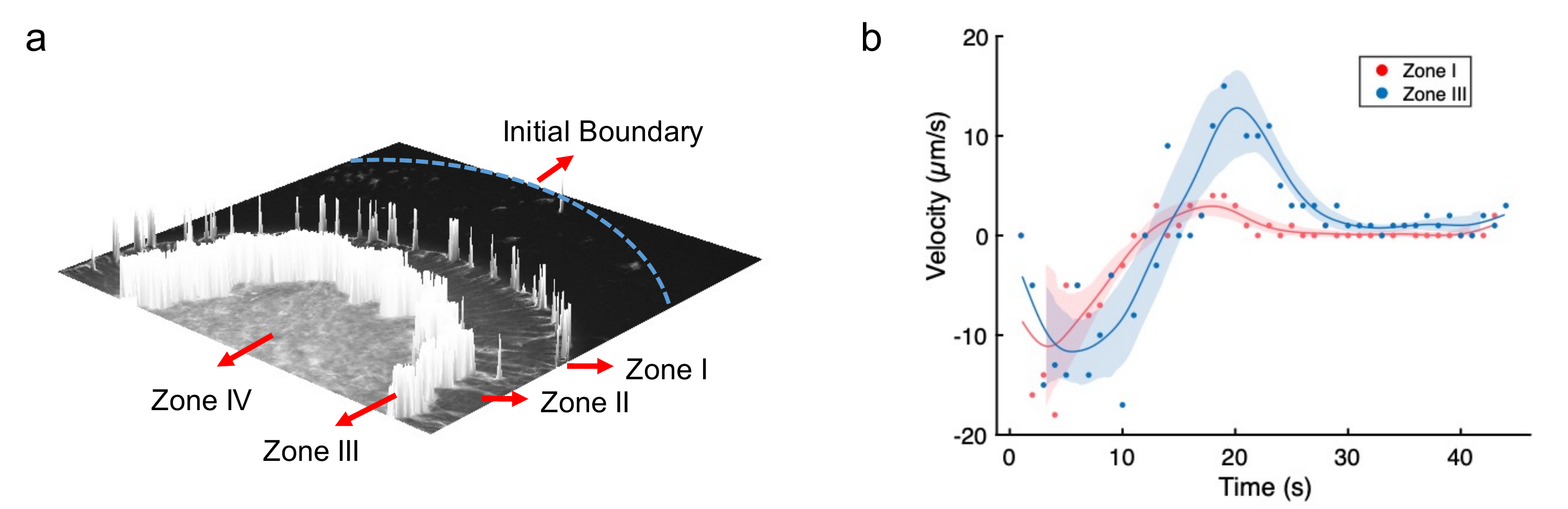}
	\caption{a) Representation of the 3D ligth profile of the pattern formed in contracting active network (by 3D Surface plot plugin of ImageJ). b) Contraction velocity measurements of the outer (zone I) and inner edges (zone III), respectively.}
	\label{sup.fig:fig1}
\end{suppfigure}

\subsection*{PIV measurements of Marangoni flow }
Quantification of the Marangoni flow by measuring the flow velocity inside the droplet was conducted with PIV measurement technique as described above.
The plot shows the cross section of the drop, where the \textit{z}-axes corresponds to the distance to the substrate surface and \textit{d} is the distance to the contact line along the substrate surface (Figure \ref{sup.fig:s2}). The blue circles and fitting line correspond to the liquid-air interface. The contact line is at origin of the plot. The arrows indicate the velocities measured in different z-planes. A strong flow is observed close to the contact line, in comparison to the flow far from the droplet ridge. The difference in velocity is around two orders of magnitude.  For better view of the active region, the inset view shows the velocity profile close to the contact line. In this region, strong Marangoni flows were developed, characterized by an inward flow at the liquid-air interface, balanced by an outward flow (capillary) close to the substrate. This configuration suggest typical Marangoni contraction of the droplet towards center of it.
\begin{suppfigure}[]
	\centering
	\includegraphics[width=\textwidth]{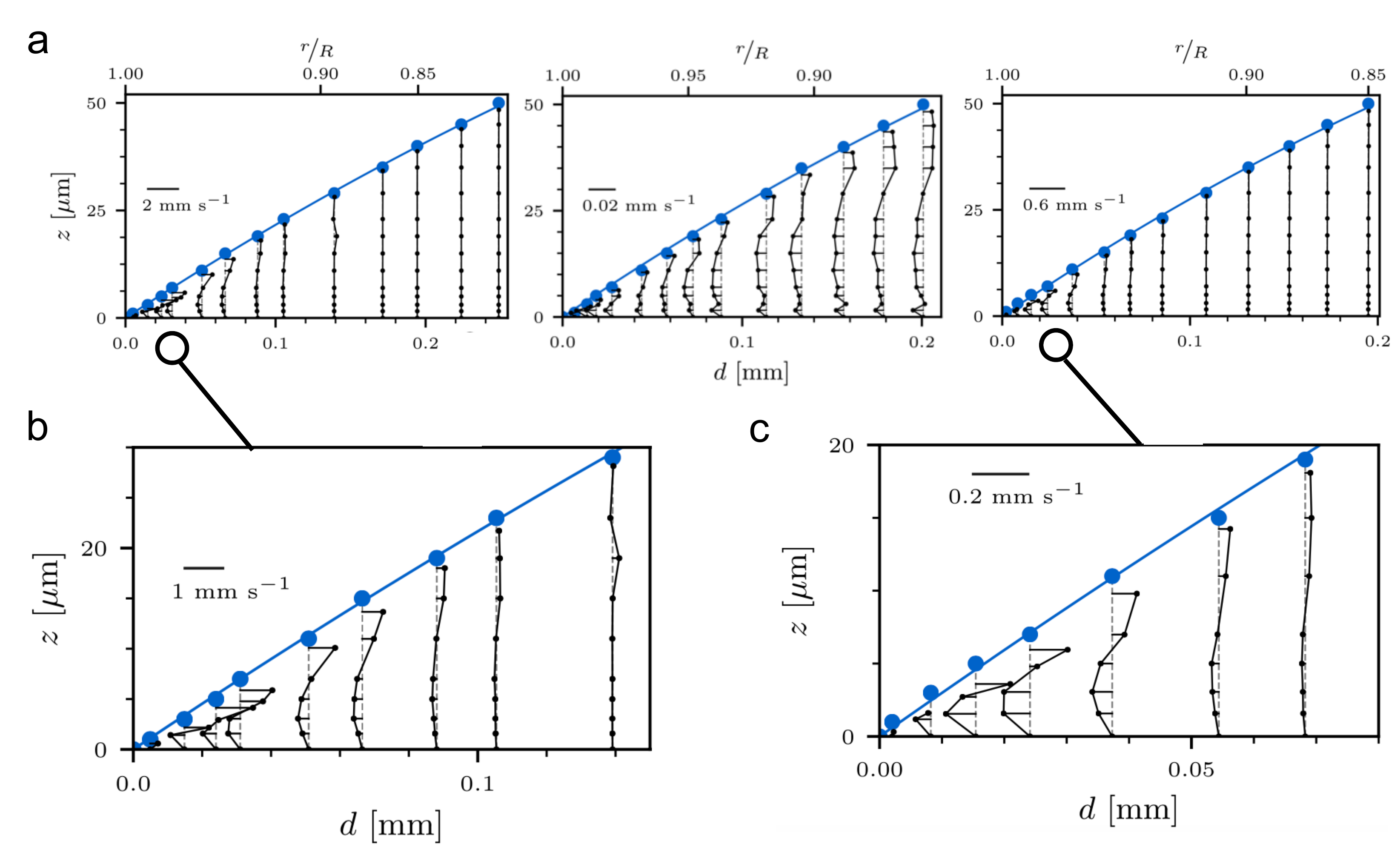}
	\caption{Velocity field measurements inside evaporating droplet made of buffer. a) Velocity field measurements for buffer droplet containing PEG with different molecular weight, namely 6, 20 and 100 kDa from left to right, respectively. b,c) Inset view of the velocity field measurements at the region of the droplet where Marangoni vortex occurs.}
	\label{sup.fig:s2}
\end{suppfigure}

\begin{suppfigure}[]
	\centering
	\includegraphics[width=0.7\textwidth]{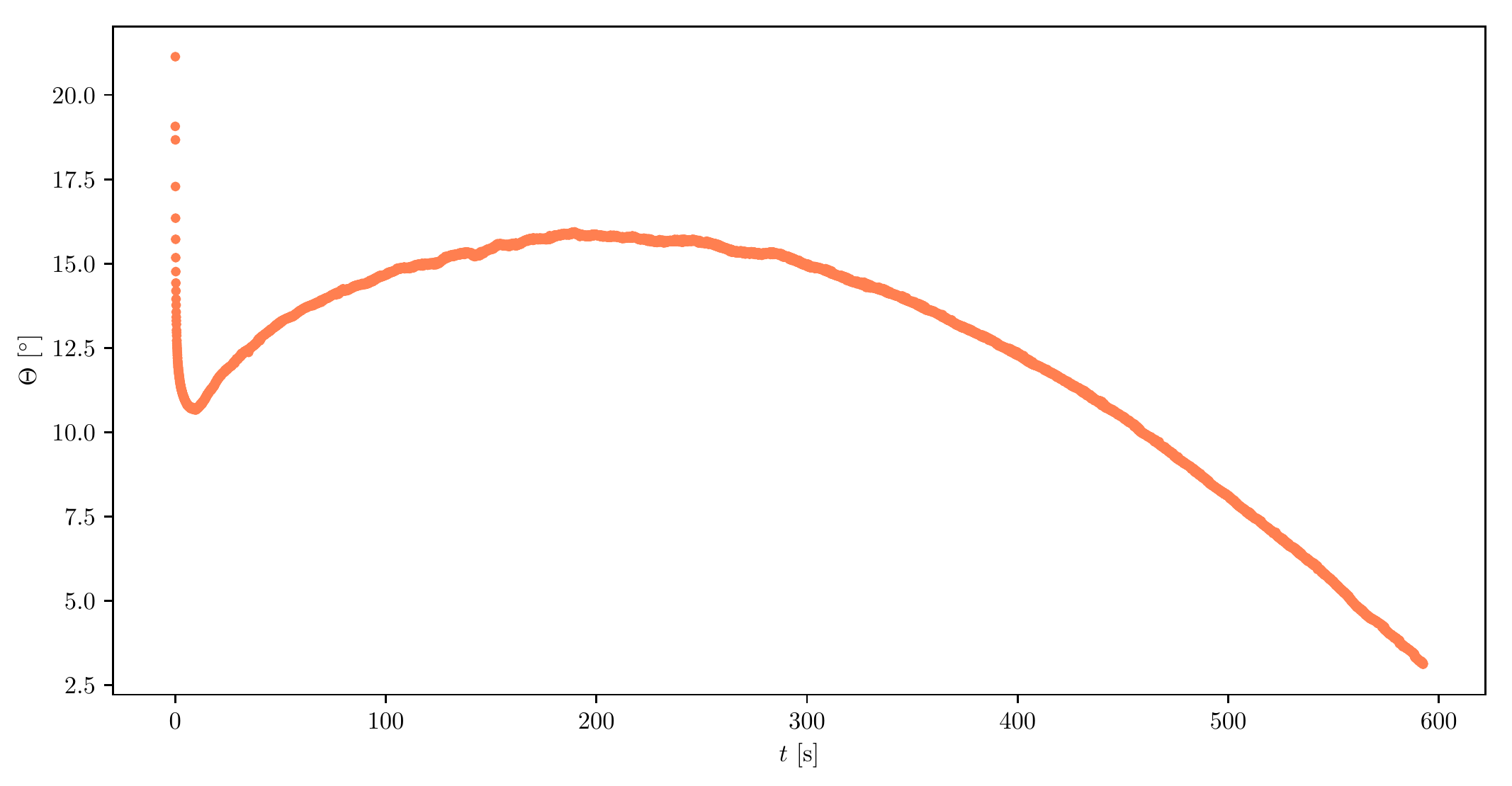}
	\caption{Measurement of the apparent contact angle of an evaporating droplet of M2B buffer containing PEG 20 kDa.}
	\label{sup.fig:s3}
\end{suppfigure}
\newpage
\subsection*{Droplet components affect contraction ratio}
We conducted experiments to understand to what extent the final contraction rate of the droplets is influenced by the presence of additional components beyond aqueous components.
We measured the contraction rate of droplets containing only PEG with different molecular weight (6, 20 and 100 KDa). We then repeated the experiment including spherical particles inside the droplet (Figure \ref{sup.fig:S4}). Comparing the contraction ratios (Figure \ref{sup.fig:S4} c-d) shows that in the absence of particles droplet contracts more compared to the case of particles inside the droplet. Exponential decay fits the data and allows to quantify  the  different contracting behaviour
\begin{suppfigure}[ht!]
	\centering
	\includegraphics[width=0.85\textwidth]{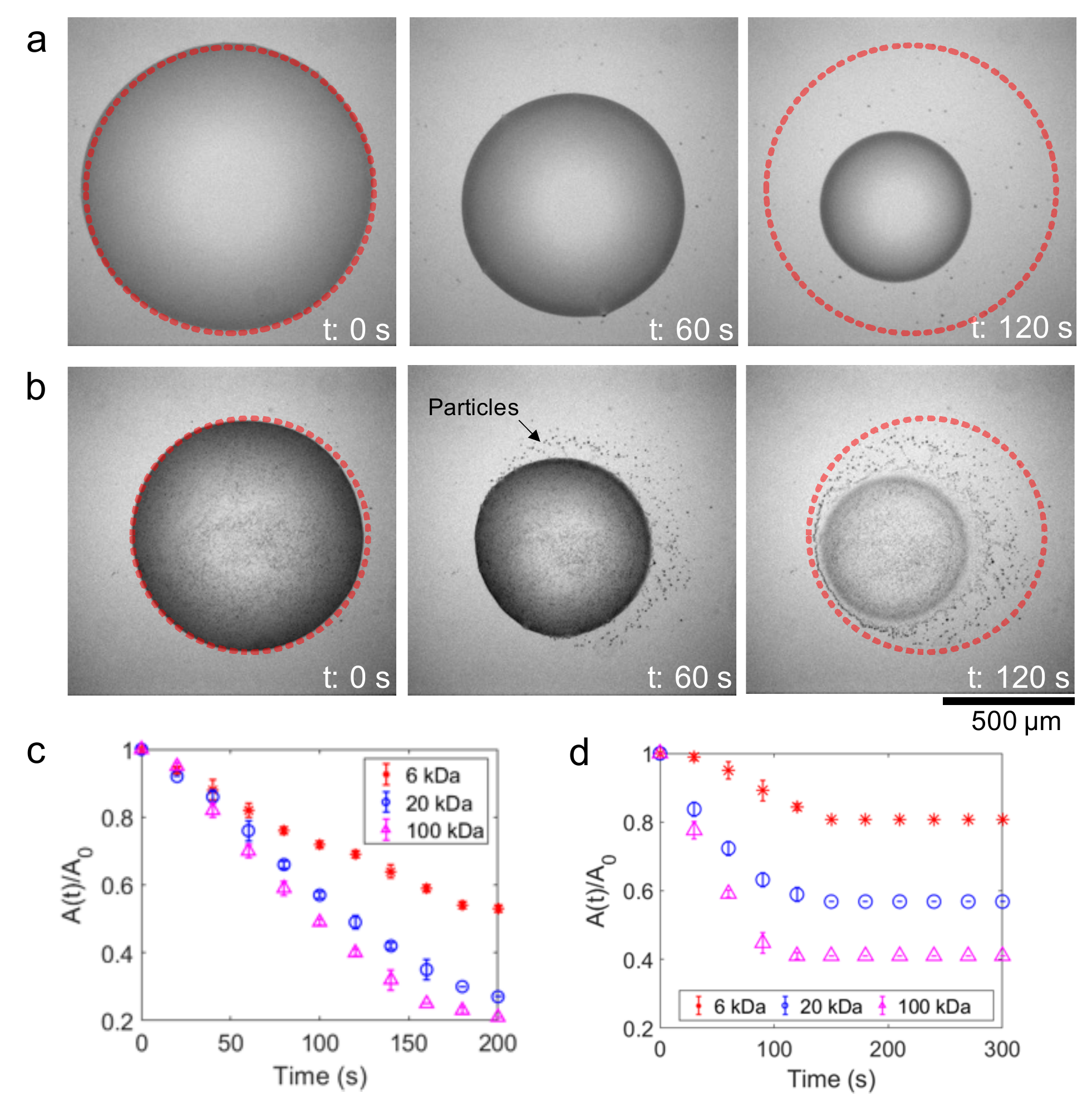}
	\caption{Comparison of contraction ratio for buffer droplet (M2B+PEG) with and without 980 nm size spherical particles. a) Time-lapse images of droplet without particles. Red dashed line represents the initial boundary of the droplet. b) Time-lapse images of the droplet containing spherical particles with a diameter of approx. 980 nm. c) Plot of the contraction ratio for buffer droplet containing PEG with different molecular weight (6, 20 and 100 KDa). The buffer droplet contracts up to 20\% of its initial surface area. d) Plot of the contraction ratio for droplet containing spherical particles, with different PEG molecular weight. The droplet contraction decayed faster and reached quickly a plateau. }
	\label{sup.fig:S4}
\end{suppfigure}
\newpage
\begin{suppfigure}[]
	\centering
	\includegraphics[width=\textwidth]{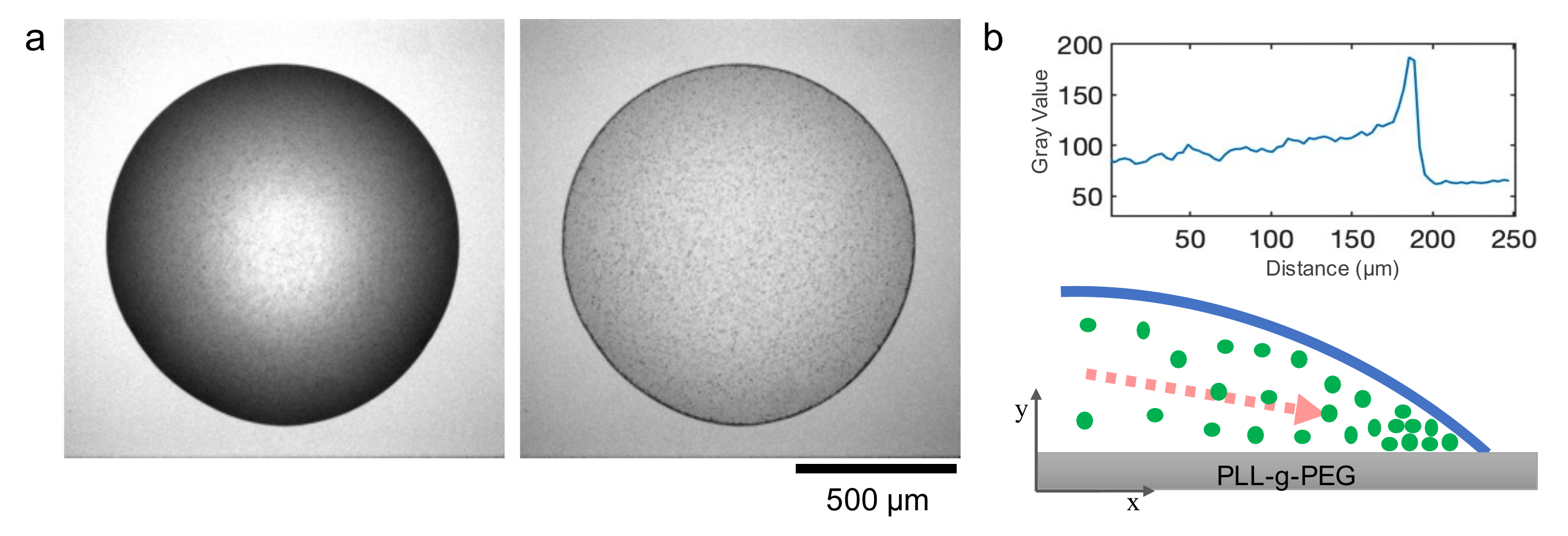}
	\caption{Formation of coffee ring by spherical particles on functionalized glass substrate. a) Wet and dried stages of an evaporating droplet containing buffer solution and spherical particles (980 nm in diamater). The coffee ring can be easily seen at the contact line of the droplet. b) Grey value intensity measurement at the contact line, showing the accumulation of particles. }
	\label{sup.fig:S5}
\end{suppfigure}

\begin{table}\centering
	\caption{Surface tension measurements of M2B buffer plus PEG.}
	\begin{tabular}{lr} 
		Composition & Surface tension $\gamma_{LV}$ (mN/m) \\
		Water  & 72 + 0.43 \\
		PEG 5\%  & 62.17 + 0.14 \\
		PEG 1\%   & 62.95 + 0.22 \\
		PEG 0.5\%   & 63.09 + 0.17 \\
		M2B 5\%   & 73.26 + 0.20 \\
		M2B 1\%   & 72.64 + 0.30 \\
		M2B 2.5\%+PEG 2.5\%  & 61.40 +0.16 \\
		M2B 2.5\%+PEG 0.25\%  & 61.85 + 0.14 \\
	\end{tabular}
	\label{table:1}
\end{table}

\end{document}